\documentclass[aps,pr,twocolumn]{revtex4}
\makeatletter

\newcommand{\Rmnum}[1]{\expandafter\@slowromancap\romannumeral #1@}
\makeatother
\usepackage{graphicx}
\usepackage{dcolumn}
\usepackage{bm}
\usepackage{CJK}
\usepackage{color}
\usepackage{amsfonts}
\usepackage{psfrag}
\usepackage{wrapfig}
\usepackage{subfigure}
\usepackage{makeidx}
\usepackage{multirow}
\usepackage{epsf}
\usepackage[colorlinks,linkcolor=red,anchorcolor=red,citecolor=blue,urlcolor=blue]{hyperref}
\usepackage{amsmath}
\usepackage{cases}
\usepackage{bm}

\begin{document}
\begin{CJK}{GBK}{song}
\title{Modulation instability and non-degenerate Akhmediev breathers of Manakov equations}
\author{Chong Liu$^{1,2,3,4}$}\email{chongliu@nwu.edu.cn}
\author{Shao-Chun Chen$^{1}$}
\author{Xiankun Yao$^{1,3}$}\email{yaoxk@nwu.edu.cn}
\author{Nail Akhmediev$^{2}$}\email{Nail.Akhmediev@anu.edu.au}
\address{$^1$School of Physics, Northwest University, Xi'an 710127, China}
\address{$^2$Optical Sciences Group, Research School of Physics and Engineering, The Australian National University, Canberra, ACT 2600, Australia}
\address{$^3$Shaanxi Key Laboratory for Theoretical Physics Frontiers, Xi'an 710127, China}
\address{$^4$NSFC-SPTP Peng Huanwu Center for Fundamental Theory, Xi'an 710127, China}
\begin{abstract}
We reveal a new class of \textit{non-degenerate} Akhmediev breather (AB) solutions of Manakov equations that only exist in the focusing case.
Based on exact solutions, we present the existence diagram of
such excitations on the frequency-wavenumber plane.
Conventional single-frequency modulation instability leads to simultaneous excitation of three
ABs with two of them being non-degenerate.
\end{abstract}

\maketitle
Generation of complex wave patterns out of simple ones is one of the basic phenomena in physics. A quintessential example is the modulation instability (MI) discovered back in 1960-ies \cite{Bespalov,BF}. Despite being known for decades, it remains the subject of high interest today \cite{Solli07,Solli12,Nguyen17,Chen20,Leykam21}.  MI is closely connected to emergence of rogue waves \cite{RW1,RW2,RW3,RW4}. It is directly related to recurrence phenomena \cite{FPU,Recurrence0,Recurrence1,Recurrence2,Recurrence3,MC1,MC2}. MI leads to breather formation \cite{TMP86,LA2021,JETP88,Erkintalo,Gelash,Trillo}. It is responsible for oscillatory structures expanding with finite speed \cite{El,Biondini,Randoux}, for creation of bound states in soliton gas \cite{Gelash2019}, for supercontinuum generation in optical fibres \cite{SCG0,SCG1,SCG2}, and even for the inception of turbulence \cite{Turbulence}.
Modern equipment allows to observe multiple growth-decay cycles induced by the MI  \cite{Recurrence2,Recurrence3,TMP86,LA2021,JETP88,Erkintalo}.
Such MI scenario leads to the Fermi-Pasta-Ulam recurrence \cite{FPU,Recurrence2,Recurrence3} and to the supercontinuum generation \cite{SCG0,SCG1,SCG2}. In systems governed by the nonlinear Schr\"odinger equation (NLSE), a full-scale growth-decay evolution can be described by the exact `Akhmediev breather' (AB) solutions \cite{TMP86}. More complex structures can be described by the superposition of several ABs (i.e., multi-ABs) \cite{JETP88}.
These solutions also describe the so-called higher-order MI which has been observed experimentally both in optics \cite{Erkintalo} and in hydrodynamics \cite{Chabchoub}.

These phenomena are turning even more complex when studying MI beyond the NLSE approximation. Indeed, the scalar NLSE describes the nonlinear dynamics of only one wave component. In reality, the nonlinear interaction of several wave components is common in optical fibres \cite{OF}, in two-component Bose-Einstein condensates \cite{BEC}, and in the case of two-directional ocean waves (so called crossing seas) \cite{F}.
The mathematical model that describes such interaction is commonly based on Manakov equations \cite{MM}. The interaction between the two wave components in the focusing case results in more complex breather dynamics \cite{VB1,VBCK,F1,F2,Z1,VNLSE1,VNLSE2,Liu2021}.
In the defocusing case, these equations admit dark rogue waves \cite{Vobservation1,Vobservation2}.
In this work, we found a new class of AB solutions of Manakov equations. They are \textit{non-degenerate} in the sense that the two AB components have unequal individual  eigenvalues. This class only exists in the focusing regime of the two-component vector fields. We derive the exact analytical form of these solutions, construct their existence diagram and show that the conventional single frequency MI leads to the simultaneous excitation of three ABs with the two of them that belong to the new class.

We consider the Manakov equations as follows:
\begin{eqnarray}\label{eqmanakov}
\begin{split}
i\frac{\partial\psi_{1}}{\partial t}+\frac{1}{2}\frac{\partial_2\psi_{1}}{\partial x^2}+\sigma(\bm|\psi_1\bm|^2+\bm|\psi_{2}\bm|^2)\psi_{1}&=0,\\
i\frac{\partial\psi_{2}}{\partial t}+\frac{1}{2}\frac{\partial_2\psi_{2}}{\partial x^2}+\sigma(\bm|\psi_1\bm|^2+\bm|\psi_{2}\bm|^2)\psi_{2}&=0,
\end{split}
\end{eqnarray}
where $\psi_{j}(t,x)$ are the two ($j=1,2$) nonlinearly coupled components of the vector wave field.
The physical meaning of independent variables $x$ and $t$ depends on a particular physical problem of interest. In optics, $t$ is commonly a normalised distance along the fibre while $x$ is the normalised time in a frame moving with group velocity \cite{OF}. In the case of  a condensate in quantum liquids, $t$ is time while $x$ is the spatial coordinate \cite{BEC}.

We start with the fundamental AB solution of Eqs.(\ref{eqmanakov}). It is valid in both the focusing (positive $\sigma$) and the defocusing (negative $\sigma$) regimes. Using a Darboux transformation scheme for vector NLSE
\cite{VNLSE2,Liu2021} and starting with the vector plane wave solution of Eqs. (\ref{eqmanakov})
\begin{equation}
\psi_{0j}=a_j \exp\{ i [{\beta_j}x + (\sigma\sum_{j=1}^{2}a_j^2- \beta_j^2/2 ) t ] \},\label{eqpw}
\end{equation}
as a seed, at the first step, we find:
\begin{equation}
\psi_{j}=\psi_{0j}\left[\frac{\cosh(\bm{\Gamma}+i\gamma_j)e^{i\eta_{1j}}+\varpi \cos(\bm{\Omega}-i\epsilon_j)e^{i\eta_{2j}}}{\cosh\bm{\Gamma}+\varpi \cos\bm{\Omega}}\right].\label{eqb}
\end{equation}
Parameters $a_j$ and $\beta_j$ in (\ref{eqpw}) are the amplitudes and wavenumbers of the two plane wave components respectively while the scalar arguments $\bm{\Gamma}$ and $\bm{\Omega}$ in (\ref{eqb}) are:
\begin{eqnarray}
\bm{\Gamma}=\omega \bm{\chi}_i\bm t,~
\bm{\Omega}=\omega \left[\bm x+ (\bm{\chi}_r+\frac{1}{2}\omega )\bm t \right]+\arg\frac{2\bm{\chi}_i}{2\bm{\chi}_i-i\omega}.
\end{eqnarray}
Here $\bm{x}=x-x_{01}$, $\bm{t}=t-t_{01}$ are shifted spatial and time variables respectively with  $x_{01}$ and $t_{01}$ being responsible for the spatial and temporal position of the breather.
Other notations in (\ref{eqb}) are:
\begin{eqnarray}
\eta_{1j}&=&\frac{\gamma_{1j}+\gamma_{2j}}{2},~~\eta_{2j}=\arg\frac{\bm{\chi}^*+\beta_j}{\bm{\chi}+\beta_j+\omega}, \\
\gamma_{j}&=&\frac{\gamma_{1j}-\gamma_{2j}}{2},~~\varpi= \Big|\frac{2\bm{\chi}_i}{2\bm{\chi}_i+i\omega} \Big|,\\
\gamma_{1j}&=&\arg\frac{\bm{\chi}^*+\beta_j}{\bm{\chi}+\beta_j},~~
\gamma_{2j}=\arg\frac{\bm{\chi}^*+\beta_j+\omega}{\bm{\chi}+\beta_j+\omega},\\
\epsilon_j&=&\log\left(\frac{(\bm{\chi}^*+\beta_j)(\bm{\chi}+\beta_j)}{(\bm{\chi}+\beta_j+\omega)(\bm{\chi}^*+\beta_j+\omega)}\right)^{1/2}.
\end{eqnarray}

An important parameter of the breather is its eigenvalue $\bm{\chi}\equiv\bm{\chi}(\sigma, a_j, \beta_j, \omega)$ with its real $\bm{\chi}_r$ and imaginary $\bm{\chi}_i$ parts. The eigenvalue satisfies the following relation:
\begin{equation}
1+\sigma\sum_{j=1}^{2}\frac{a_j^2}{(\bm{\chi}+\beta_j)(\bm{\chi}+\omega+\beta_j)}=0.\label{eqchi}
\end{equation}
Thus, the AB solution (\ref{eqb}) depends on the following parameters: the background amplitudes $a_j$ and wave numbers $\beta_j$, the modulation frequency $\omega$, and the sign of the nonlinearity $\sigma$. Some of them can be eliminated using a Galilean transformation. Without loss of generality, we can set ${\beta_1}=-{\beta_2}=\beta$.
Also, the amplitudes of the two plane waves can be taken to be equal $a_1=a_2=a$.

As the AB solution (\ref{eqb}) represents the full cycle of modulation instability, it
grows out of the plane wave (\ref{eqpw}) weakly modulated  with frequency $\omega$.
The growth rate of the instability is $G=|\omega \bm{\chi}_i|$. It depends on the frequency $\omega$ and the eigenvalue $\bm{\chi}$. A drastic difference from the case of ABs of a scalar  NLSE is the presence of several complex eigenvalues $\bm{\chi}=\bm{\chi}_r+i\bm{\chi}_i$ which are the solutions of Eq. (\ref{eqchi}).
Importantly, the sign of $\bm{\chi}_i$ has no effect on the AB solution while the sign of $\bm{\chi}_r$ determines the spatiotemporal distribution of the AB.
In order to show this, we use
the notation $\bm{\tilde{\chi}}=\bm{\chi}+\omega/2$.
Then the explicit expressions for $\bm{\tilde{\chi}}$ obtained using Eq. (\ref{eqchi}) are:
\begin{eqnarray}
\bm{\tilde{\chi}}_{1}&=& (\bm{\mu}-\sqrt{\bm{\nu}})^{1/2},~~
\bm{\tilde{\chi}}_{2}= -(\bm{\mu}-\sqrt{\bm{\nu}})^{1/2},\label{eqchi1}\\
\bm{\tilde{\chi}}_{3}&=& (\bm{\mu}+\sqrt{\bm{\nu}})^{1/2},~~
\bm{\tilde{\chi}}_{4}=-(\bm{\mu}+\sqrt{\bm{\nu}})^{1/2},\label{eqchi2}
\end{eqnarray}
where $\bm{\nu}=a^4-4\sigma a^2\beta^2+\omega^2\beta^2$, and $\bm{\mu}=\beta^2-\sigma a^2+\omega^2/4$.

In the case $\sigma=-1$, $\bm{\chi}_3\in \mathbb{R}$, and $\bm{\chi}_4\in \mathbb{R}$ while $\bm{\chi}_1=\bm{\chi}_2^*$ ($*$ denotes the complex conjugate). Thus, the defocusing regime only admits degenerate AB for every modulation frequency $\omega$. We omit this case from our analysis. On the other hand, in the focusing regime ($\sigma=1$), two different cases follow from Eqs. (\ref{eqchi1}) and (\ref{eqchi2}): $\bm{\nu}\geq0$ and $\bm{\nu}<0$.
These conditions separate the solutions to the classes of degenerate and non-degenerate vector ABs.

\begin{figure}[htb]
\centering
\includegraphics[width=85mm]{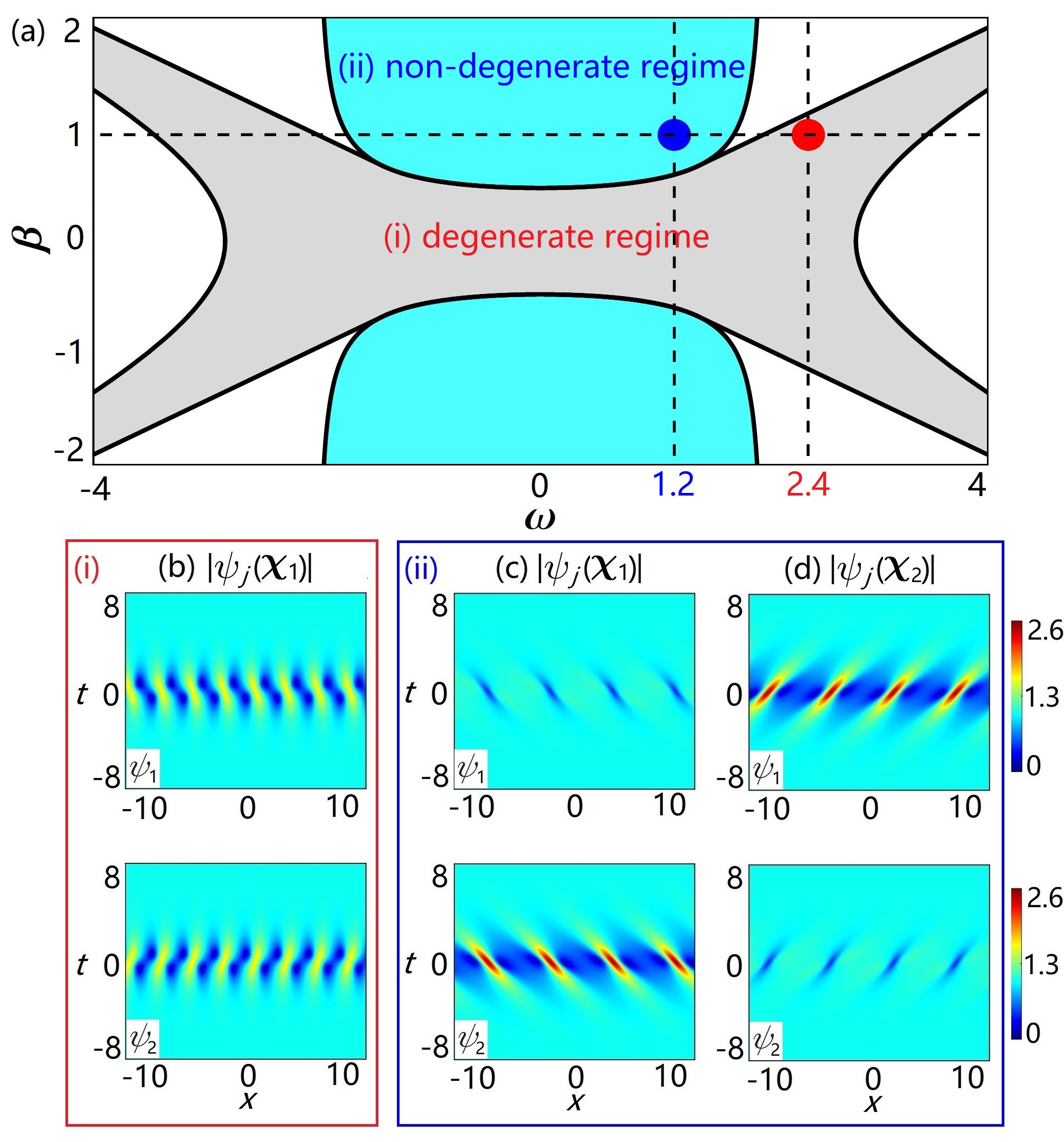}
\caption{(a) The existence diagram of degenerate and non-degenerate ABs on the ($\omega-\beta$) plane in the focusing case ($\sigma=1$), constructed from the analysis of eigenvalues (\ref{eqchi1}) and (\ref{eqchi2}).
(b) The amplitude profiles $|\psi_j(\bm{\chi}_{1})|$ of the degenerate vector AB solution (\ref{eqb}) with $\omega=2.4$ and $\beta=1$ (large red dot in (a)). (c)  The amplitude profiles of the non-degenerate vector AB solution with $\omega=1.2$ and $\beta=1$ (large blue dot in (a)). (d) Inverted non-degenerate vector AB solution $|\psi_j(\bm{\chi}_{2})|$ for the same parameters as in (c).
The background $a=1$ while $x_{01}=0$ and $t_{01}=0$.}\label{f1}
\end{figure}

When $\bm{\nu}\geq0$ (or $\beta^2\leq a^4/(4a^2-\omega^2)$), we have $\bm{\chi}_{1}=\bm{\chi}_{2}^*$, $\bm{\chi}_{3}=\bm{\chi}_{4}^*$. Only the eigenvalues $\bm{\chi}_{1}$ and $\bm{\chi}_{2}$ can be used in the AB solutions (\ref{eqb}). Indeed, when $\beta=0$, the solution becomes decoupled.
This results in $\bm{\chi}_{1}=\bm{\chi}_{2}^*=\sqrt{\omega^2/4-2a^2}$, and $\bm{\chi}_{3}=\bm{\chi}_{4}=0$. This means that the two components of coupled wave fields are equal, and the solution (\ref{eqb}) must coincide with the scalar NLSE AB solution.  Thus, when $\bm{\nu}\geq0$, the eigenvalues $\bm{\chi}_{3}$, $\bm{\chi}_{4}$ must be eliminated, and only $\bm{\chi}_{1}$, $\bm{\chi}_{2}$ remain valid eigenvalues. As $\bm{\chi}_{1}=\bm{\chi}_{2}^*$, the two components are: $\psi_j(\bm{\chi}_{1})=\psi_j(\bm{\chi}_{2})$.
Such solutions belong to the class of \textit{degenerate} ABs.
An example of such degenerate ABs is shown in Fig. \ref{f1}(b).

On the other hand, when $\bm{\nu}<0$, (or $\beta^2>a^4/(4a^2-\omega^2)$), we have $\bm{\chi}_{1}=\bm{\chi}_{3}^*$, $\bm{\chi}_{2}=\bm{\chi}_{4}^*$. Then,
$\bm{\chi}_{1i}=-\bm{\chi}_{2i}$ and $\bm{\chi}_{1r}\neq \bm{\chi}_{2r}$.
This means that there are two valid eigenvalues ($\bm{\chi}_{1}$, $\bm{\chi}_{2}$) with different real parts ($\bm{\chi}_{1r}\neq\bm{\chi}_{2r}$).
In this case, $\psi_j(\bm{\chi}_{1})\neq\psi_j(\bm{\chi}_{2})$.
As a result, there are two ABs with different wave profiles ($\bm{\chi}_{1r}\neq \bm{\chi}_{2r}$) for any given background wave ($a$, $\beta$) and a fixed modulation frequency $\omega$.
The two wave profiles share the same growth rate ($\bm{\chi}_{1i}=-\bm{\chi}_{2i}$).
Thus, solutions in this class are \textit{non-degenerate}.
An example of such non-degenerate solution and its inverted version ($\psi_j(\bm{\chi}_{1})\leftrightarrow\psi_j(\bm{\chi}_{2})$) are shown in Figs. \ref{f1}(c) and \ref{f1}(d) respectively.

These the nontrivial generalisations of a scalar AB to the vector case which have not been reported previously. As can be seen from Fig. \ref{f1}(c), there is an asymmetry between the profiles of the two wave components $\psi_j(\bm{\chi}_{1})$, and $\psi_j(\bm{\chi}_{2})$.
Firstly, the amplitude profile of $|\psi_1|$ component exhibits periodic `dark' structures in $x$ while $|\psi_2|$ reveals more conventional AB pattern with `bright' peaks. Secondly, the direction of wave propagation is tilted. Despite the asymmetry, the two components of such non-degenerate ABs share the same modulation frequency and the same growth rate.
Clearly, the asymmetric solution can exist in a symmetric system.
However, the symmetry of the Manakov equations requires the existence of a solution with reversed components. Such solution is shown in Fig. \ref{f1}(d). Here, the two components reversed and inverted in $x$.

Clearly, each of these ABs can be excited individually by using the ideal initial conditions in the form of the exact solutions at any fixed (large) negative $t$. Then the evolution will follow the AB solution which is the separatrix in an infinite-dimensional phase space.
A practical question is whether they can be excited from the plane wave with a simple sinusoidal modulation. In the vector case, this problem is not as trivial as it seems.
In order to address it, we have solved the Manakov equations (\ref{eqmanakov}) numerically using the split-step Fourier method. We used, as the initial condition, the plane wave with a single-frequency modulation in each component:
\begin{equation}
\psi_{j}=(1+\varepsilon\cos\omega x)\psi_{0j},\label{eqin}
\end{equation}
where $\varepsilon$ ($\ll1$) is a small amplitude of modulation, while $\omega$ is its frequency. Linear stability analysis around the plane wave (\ref{eqpw}) shows that it is a saddle point in an infinite-dimensional phase space \cite{TMP86}. The weakly perturbed point (\ref{eqin}) is located on one of the trajectories around this saddle point. There are two strong restrictions in Eq. (\ref{eqin}). Firstly, $\epsilon$ is real. This means that the initial point is not located exactly on the separatrix and the evolution does not have to follow it. Secondly, the amplitudes of the two components in (\ref{eqin}) are the same. This imposes the symmetry on the solution that is absent in the case of ABs shown in Figs. \ref{f1}(c) and (d). However, this symmetric initial condition does excite the non-degenerate AB solutions if the frequency $\omega$ is chosen in the blue region in Fig. \ref{f1}a.

\begin{figure}[htb]
\centering
\includegraphics[width=85mm]{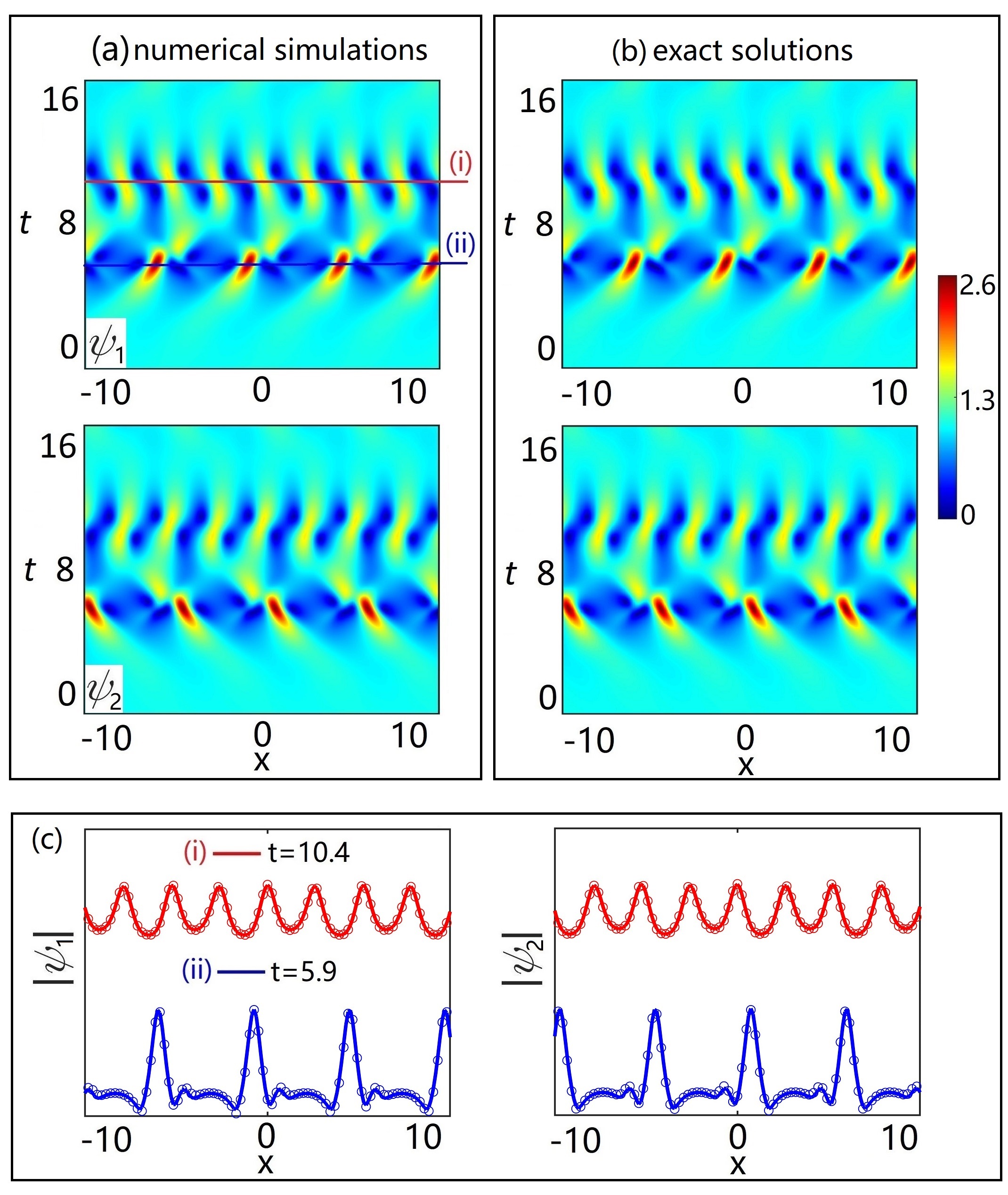}
\caption{(a) The results of numerical simulations started from the initial condition (\ref{eqin}) with $\omega=1.2$, $\beta=1$, $\varepsilon=0.01$, and $a=1$.
(b) Higher-order exact solution of Manakov equations that consists of three ABs (\ref{eqb})
with parameters $\omega_1=\omega_2=1.2$, $\omega_3=2.4$, $t_{01}=6.3$, $x_{01}=-3.917$, $t_{02}=5.7$, $x_{02}=-2.867$, $t_{03}=8.6$, and $x_{03}=-0.819$. (c) Wave profiles of the first (left) and the second (right) components at the points of maximal pulse compression in the two cycles. Solid curves correspond to numerical simulations shown in (a) while the circles correspond to the exact solution shown in (b).
} \label{f2}
\end{figure}

Figure \ref{f2}(a) shows the results of the numerical simulations. The parameters used in (\ref{eqin}) are $a=1$, $\beta=1$, $\omega=1.2$, and $\varepsilon=0.01$. The frequency $\omega=1.2$ is chosen specifically to excite the non-degenerate ABs. Nevertheless, the choice of equal components in (\ref{eqin}) leads to the simultaneous excitation of both of them. The two excited ABs are inverted copies of each other. They are excited in the form of a nonlinear superposition with the point of the maximum pulse compression in time that is located at $t=5.9$ (the point (ii) on the time axis). Therefore, the first growth-return cycle in Fig. \ref{f2}(a) becomes more complex in comparison with the elementary ABs shown in Fig. \ref{f1}(c) or (d). This first cycle ends at another saddle point which is the plane wave background with a small periodic perturbation although different from (\ref{eqin}). Due to the transverse phase shift introduced by each non-degenerate AB, the perturbation now contains the second harmonic of the initial frequency $\omega$.

The second growth-return cycle starts at this saddle point perturbed with the frequency $2\omega=2.4$. As a consequence, the second growth-return cycle involves the degenerate AB shown in Fig. \ref{f1}(a) by the thick red circle.
 Comparison of this cycle with the field profiles shown in Fig. \ref{f1}(b) shows that it indeed belongs to the class of degenerate ABs. The exponential return back to the plane wave is closer to the separatix. No further cycles appear within the timeframe shown in Fig. \ref{f2}(a).
 To summarise, the wave evolution shown in Fig. \ref{f2}(a) is a nonlinear superposition of three ABs rather than just two ones.

\begin{figure}[htb]
\centering
\includegraphics[width=85mm]{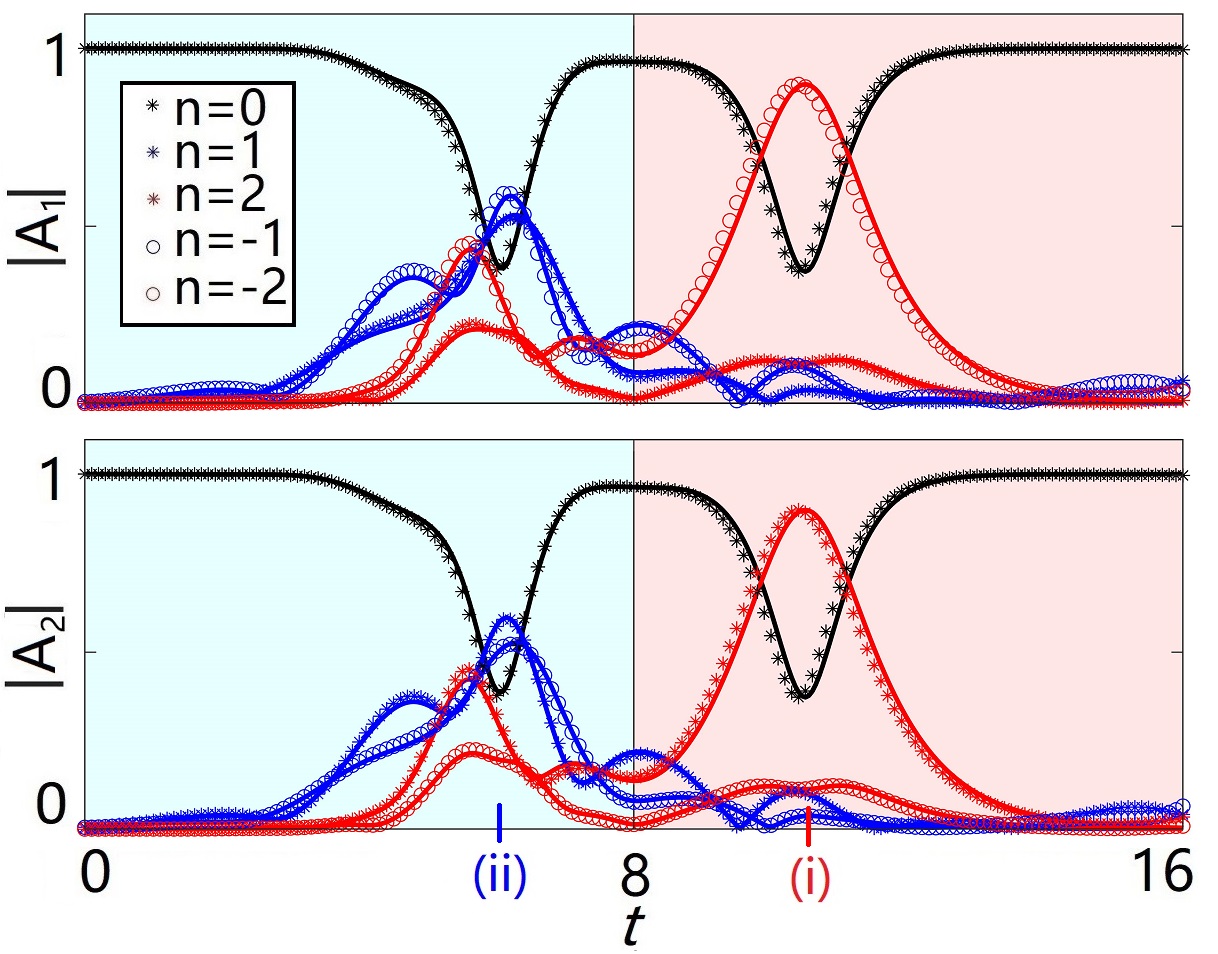}
\caption{Evolution of the lowest order Fourier spectra of the two wave components ($A_j\leftrightarrow\psi_j$) for the same data as in Fig. \ref{f2}. The result of numerical simulations are shown by solid lines while the spectra calculated from exact solution are shown by stars and circles. Black curves correspond to the central frequency ($n=0$), blue curves correspond to the first sidebands ($n=\pm1$) and the red curves correspond to the second sidebands ($n=\pm2$).}\label{f3}
\end{figure}

In order to further prove this conclusion, we constructed the exact third-order AB solution to match the results shown in Fig. \ref{f2}(a). This requires the next two iterations of the Darboux transformation scheme presented in \cite{NDS}. In this approach, the $n$th-order solution corresponds to the nonlinear superposition of $n$ first-order ABs, each associated with a different eigenvalue $\bm{\chi}$ and with individual parameters $(\omega_i, x_{0i},t_{0i})$. The spatiotemporal amplitude profile of such multi-AB depends on the relative separations of individual ABs in both $x$ and $t$.
Figure \ref{f2}(b) shows the amplitude profile of the nonlinear superposition of the three elementary ABs shown in Figs. \ref{f1}(a) and (b).
As we can see from the comparison of Figs. \ref{f2}(b) and \ref{f2}(a), the exact solution matches very well the numerical simulations confirming the qualitative description of the evolution given above.
The accuracy of matching can be further seen from the detailed comparison of the wave profiles $|\psi_1|$ and $|\psi_2|$ of the two components found in the exact solution and in numerical simulations at the time of maximum pulse compression of each cycle that are shown in Fig. \ref{f2}(c). The blue ($t=5.9$) and red ($t=10.4$) solid curves taken from numerical simulations are in excellent agreement with the exact solutions shown by the blue and red circles respectively.

Comparing the exact solution with numerical simulations, we emphasise another nontrivial point. Despite the symmetric initial condition (\ref{eqin}), the centres of the two non-degenerate ABs are located at different points: $t_{01}=6.3$, $x_{01}=-3.917$  and $t_{02}=5.7$, $x_{02}=-2.867$. Thus, the complex dynamics that starts from (\ref{eqin}) is still not symmetric.

We also compared the lowest order ($n=0, \pm1, \pm2$) spectral Fourier components calculated during the above wave evolution both in numerical simulations and for the exact solution. These results are shown in Fig. \ref{f3}. The upper panel shows the spectra of the first wave component while the lower panel shows the spectra of the second component. The  spectra of exact solutions are calculated by the method proposed in \cite{LA2021,Liu2021}. The numerical spectra follow directly from the numerical integration. There is an excellent agreement between them for all spectral components up to $n=\pm2$.
Initially, most of the energy is concentrated in the central ($n=0$) spectral component ($A_j=1$). Upon evolution, the first ($n=\pm1$) and the second ($n=\pm2$) sidebands gain power at the expense of the central component. At the point (ii) of maximal pulse compression in the first cycle, most of the energy is concentrated in the first ($n=\pm1$) spectral component (blue curves). Energy returns back to the central component after the first grow-return cycle at $t\approx 8$. Both the first and the second spectral components are present at this point. However, mostly the second harmonic is amplified to its maximum value during the second growth-return cycle. Consequently,
at the point of maximal pulse compression of the second cycle, most of the energy is concentrated in one of the second spectral components ($n=\pm2$). The asymmetry between the left and right components is related to the skewed propagation of the degenerate AB seen in Figs. \ref{f1}(b), \ref{f2}(a) and \ref{f2}(b). After the second cycle, energy returns to the central component. This is in full agreement with the near recurrence back to the plane wave that can be seen in Figs. \ref{f2}(a) and \ref{f2}(b).

The existence of the class of non-degenerate ABs shines a new light on the dynamics of ocean waves in two dimensions. In the crossing seas \cite{F}, the asymmetry once appeared may grow in one direction despite of the symmetry of all other conditions. Moreover, a small asymmetry in the wave pattern at the beginning of an instability may lead to a disproportional growth of waves in a single direction. This asymmetry adds to the unpredictability of nonlinear waves in the open ocean that has to be taken into account in  forecasting algorithms.

Considering wide range of applications of the Manakov equations, the new class of non-degenerate ABs has to be taken into account in a variety of vector nonlinear systems: ocean wave dynamics, fibre optics, plasma physics and multi-component quantum liquids.

This work is supported by National Natural Science Foundation of China (NSFC) (No. 12175178, No. 12047502, 12004309, and No. 11705145), and the Major Basic Research Program of Natural Science of Shaanxi Province
(No. 2017KCT-12 and No. 2017ZDJC-32).

\end{CJK}

\end{document}